\documentclass[%
reprint,
amsmath,amssymb,
aps,
]{revtex4-1}
\usepackage{multirow}
\usepackage{color}  
\usepackage{graphicx}
\usepackage{dcolumn}
\usepackage{bm}

\usepackage{subcaption}
\begin{document}
	
	
	\title{Theoretical investigation of  quantum capacitance in the functionalized MoS$_2$-monolayer }
	
	\author{Sruthi T, Nayana Devaraj}
	\author{Kartick Tarafder}%
	\email{karticktarafder@gmail.com}
	\affiliation{%
		Department of Physics, National Institute of Technology Karnataka, Surathkal, PO: Srinivasnagar,
		Mangalore – 575025.
	}%

	\date{\today}

\begin{abstract}
In this work, we investigated the electronic structure and the quantum capacitance of the functionalized MoS$_2$ monolayer. The functionalizations have been done by using different ad-atom adsorption on Mo$S_2$ monolayer. Density functional theory calculations are performed to obtain an accurate electronic structure of ad-atom doped MoS$_2$ monolayer with a varying degree of doping concentration. The quantum capacitance of the systems was subsequently estimated.  A marked quantum capacitance above 200 $\mu$F/cm$^2$ has been observed. Our calculations show that the quantum capacitance of MoS$_2$ monolayer is significantly enhanced with substitutional doping of Mo with transition metal ad-atoms. The microscopic origin of such enhancement in quantum capacitance in this system has been analyzed. Our DFT-based calculation shows that generation of new electronic states at the proximity of the band-edge and the shift of Fermi level caused by the ad-atom adsorption results in a very high quantum capacitance in the system.
 
\end{abstract}
%
\maketitle
%

\section*{Introduction}
The creation of adequate energy from sustainable energy sources, bypassing the utilization of fossil fuels, is one of the greatest challenges to stifle the weather change. Large-scale generation of green energy from sustainable power sources is, therefore, highly essential in the current situation. On the other hand, technology needs to be developed that can efficiently convert and store the generated energy. Batteries, fuel cells, and supercapacitors are the essential technological devices that help to convert and store energy. Of these, supercapacitors have received the most attention in recent times due to its high power density, long lifetime, good stability, and their possible applications over a wide temperature range  \cite{xu2019adsorption,simon2010materials}. Supercapacitors can temporarily store a large amount of electrical energy and release it when needed. However, their low energy density and high production cost are a hindrance to their progress \cite{wang2012review}. It is essential to overcome these drawbacks to expand the use of supercapacitors. 
Intense research is going on to overcome the obstacle - low energy density- by changing the electrode material of supercapacitors. The most commonly considered electrodes are carbon-based materials because of their high specific surface area, good electrical conductivity, and good stability \cite{huang2012graphene,hirunsit2016electronic}. One of the essential requirements for a supercapacitor is the electrode material should possess a large specific area with significantly high ion density. Two-dimensional materials that are in general, possess a large specific area, therefore, could be the best alternative for supercapacitor electrode applications. 
Graphene, an atom-thick 2D material in which carbon atoms are arranged in a honeycomb lattice, has remarkable electrical and mechanical properties. It has been reported recently that graphene can also be used as a suitable building block for supercapacitors \cite{choi2011facilitated}. Studies have been carried out by considering graphene, chemically modified graphene, and graphene-based composites as supercapacitor electrodes \cite{stoller2008graphene,zhu2011carbon,yang2013liquid,kim2011high}. However, the poor scalability \cite{reina2009large,kosynkin2009longitudinal} hinders its large scale application, which necessitated to search for other suitable two dimensional materials for supercapacitor electrodes. Transition metal dichalcogenides can take an active part in designing such materials. 
MoS$_2$ belongs to this class of materials is already well-known to the materials science community, has already been proven to be very suitable for applications in many fields including catalysis, lithium-ion batteries, phototransistors due to their unique morphology, excellent electrical and mechanical properties \cite{merki2011recent,chang2011situ,li2011mos2,yin2012single}. 
Bulk MoS$_2$ consists of S-Mo-S units, which are held together by weak van der Waals interactions \cite{bromley1972band}. Therefore, using physical or chemical methods, single or few layers of MoS$_2$ can be easily exfoliated from bulk MoS$_2$ \cite{ramakrishna2010mos2,chodankar2020graphene}. The excellent electrical and electrochemical properties, mechanical flexibility, versatile electronic states and good environmental characteristics of MoS$_2$ can also be used as a suitable candidate for supercapacitor electrode applications \cite{hwang2011mos2}\cite{lin2014atomic}\cite{wang2014atomic}. 

In 2007, Soon and Lohz investigated the double-layer electrochemical capacitance of nano-walled MoS$_2$ film using electrochemical impedance spectroscopy and have shown that edge-aligned MoS$_2$ thin films can act as a supercapacitor at various current frequencies, which can be compared to a carbon nanotube array electrode. They also have found that ion diffusion at slow scanning speeds results in Faradaic capacitance, which greatly improves the capacitance \cite{soon2007electrochemical}. In 2013 Ma et.al reported a method to synthesize a polypyrrole/MoS$_2$(PPy/MoS$_2$) nano-composite as an innovative electrode material for efficient supercapacitors \cite{ma2013situ}. Another method adopted to develop better supercapacitors is to increase capacitance by doping of different atoms to the electrode materials. Xu et.al explored the change in quantum capacitance due to doping and co-doping on graphene-based electrodes using first principle methods \cite{xu2019improving}. They have found that the N/S and N/P co-doped graphene with vacancy defects are suitable for asymmetric supercapacitors. In one of our previous work, we have observed very high quantum capacitance in the functionalized graphene modified with ad-atoms \cite{sruthi2019route}. So it is clear from the available literature that with proper functionalization, MoS$_2$ can be made a suitable material for the super-capacitor electrode applications. Therefore, in this work, we have functionalized the MoS$_2$ with different ad-atoms to modify the quantum capacitance of supercapacitors having a monolayer MoS$_2$ electrode and investigated the microscopic origin of these changes. 

\section*{Methodology}
Structural optimization and subsequently the electronic structure information of different functionalized systems are obtained by using first-principle density functional theory calculations as implemented in the Vienna Ab-initio Simulation Package   VASP)\cite{kresse1996efficient,kresse1996efficiency}. The Generalized-gradient approximation with the Perdew–Burke–Ernzerhof (PBE) parameterization was used to describe the exchange-correlation energy \cite{perdew1996generalized}. A very high value of the energy cut-off ($\textgreater$400eV) was taken into consideration to obtain accurate results. To study the impact of various ad-atom substitution on the quantum capacitance, calculations were done utilizing 3$\times$3$\times$1 supercells of MoS$_2$ unit cell, having nine Mo atoms and eighteen sulfur atoms. Single Mo or S atomic sites were selected for substitutional doping. The vacancy defected configurations were studied on the same supercell of MoS$_2$ with varying vacancy concentration \cite{feng2018influence}. A large void space (height$\textgreater$10\AA) was considered along the out of the plane direction of layer MoS$_2$ unit cells to prevent the interaction with its periodic images. We have used a 6$\times$6$\times$1 Monkhorst-Pack grids to sample the Brillouin zone for geometrical optimization with $10^{-6}$ H total energy tolerance for convergence. A denser 24$\times$24$\times$1 Monkhorst-Pack grids were used for the precise extraction of electron density of states D(E) and atom projected density of states (PDOS).

 The total capacitance, $C_T$ of an electrical double-layer capacitor (EDLC) is expressed as 
\begin{equation}
\frac{1}{C_T} = \frac{1}{C_Q} + \frac{1}{C_D}
\end{equation}
where $C_Q$ and $C_D$ are quantum capacitance and double-layer capacitance. The quantum capacitance of materials is defined as the rate of variation of excessive charges (ions) over the change in applied potential \cite{mousavi2015first}. So it is directly related to the electronic energy configuration of the electrode materials and can be defined as the derivative of the net excess charge on the substrate/electrode with respect to electrostatic potential. 
ie,
\begin{equation}
C_Q = \frac{dQ}{d\phi}
\end{equation}
where Q is the excessive charge on the electrode and $\phi$ is the chemical potential. The total charge is in proportion to the weighted sum of the electronic DOS up to the Fermi level, $E_F$. Due to an applied potential,  the chemical potential will be shifted, the excessive charge on the electrode (Q) then can be expressed by an integral term associated with the electronic density of state D(E) and the Fermi-Dirac distribution function f(E) as
\begin{equation}
Q = e\int_{-\infty}^{+\infty}D(E)[f(E)-f(E-\phi)]dE
\end{equation}
Therefore, when the density of states (DOS) is known, the $C_Q$ of a channel at a finite temperature T can be calculated as
\begin{equation}
C_Q = \frac{dQ}{d\phi} = \frac{e^2}{4kT}\int_{-\infty}^{+\infty}D(E) sech^2 \frac{E-e\phi}{2kT} dE
\label{EqnCq}
\end{equation}
Here, $\phi$, e, and k are chemical potential, charge of an electron, and the Boltzmann constant respectively.

\section*{Results and Discussions}
\begin{figure} [t!]
 	\centering
	\includegraphics[width=0.80\linewidth]{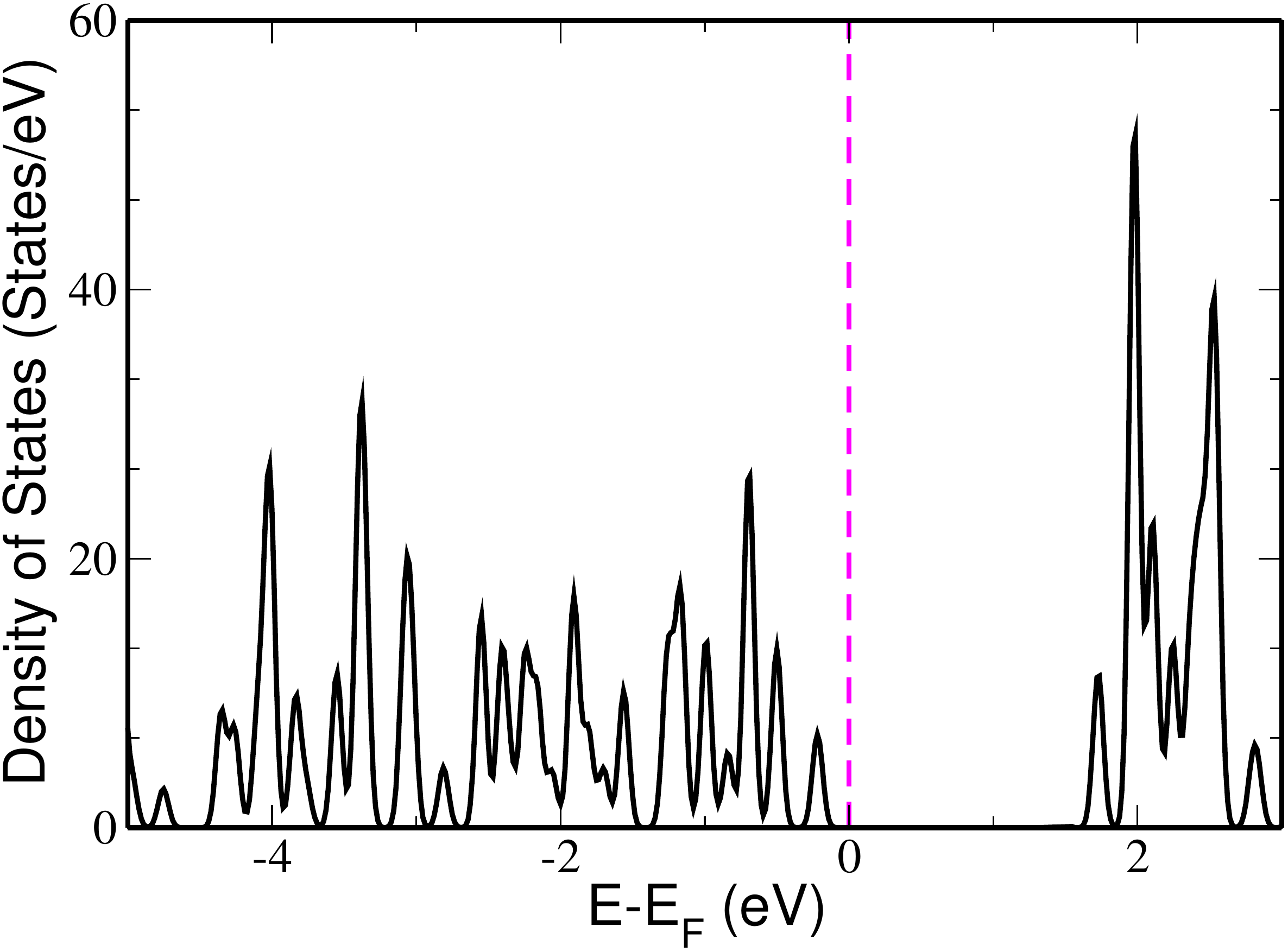}
	\caption{(color online) Density of states of MoS$_2$. Vertical majenta dashed line is the Fermi energy set at E=0.}
	\label{Fig_DOS-MoS2}
\end{figure}

The equation ($\ref{EqnCq}$) shows the role of the DOS present near the Fermi  energy. Since the monolayer  MoS$_2$ is a direct band semiconductor \cite{mak2010atomically}, in which no states are present near Fermi level (Fig.\ref{Fig_DOS-MoS2}), it gives zero quantum capacitance. The electronic structure of a material can be modified using ad-atom doping. If the change in electronic structure could accumulate states near Fermi level then the quantum capacitance can be generated in the monolayer MoS$_2$. 

\subsection*{Substitutional doping effect of Monolayer MoS$_2$}

\begin{figure} [t!]
 	\centering
	\includegraphics[width=0.8\linewidth]{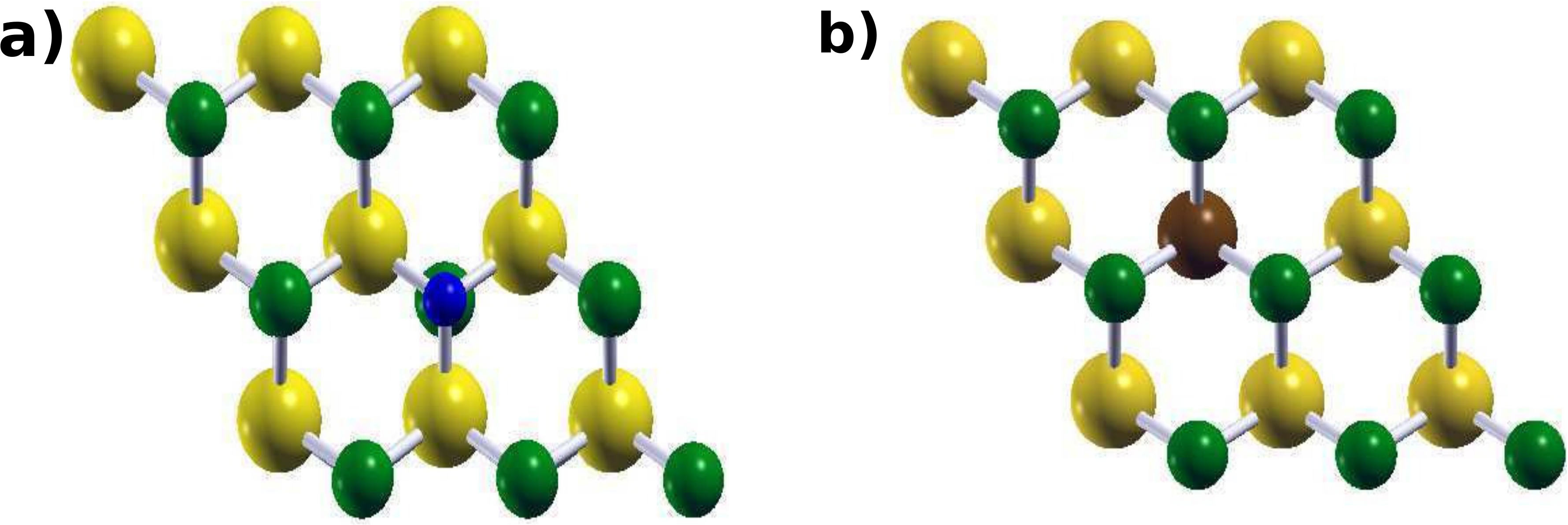}
	\caption{(color online)Structure of functionalized MoS$_2$. Green balls and yellow balls represent S and Mo respectively. a)Structure of S substituted MoS$_2$. Blue ball represent the atom which substitute S. b) Structure of Mo substituted MoS$_2$. Brown ball represent the atom which substitute Mo.}
	\label{Fig_subst-MoS2}
\end{figure}

The substitutional doping on MoS$_2$ can be done by selecting atoms from groups in the periodic table which are nearer to Mo and S atom. Hence we have investigated the change in quantum capacitance of MoS$_2$ by substituting S atoms with group five atoms as well as atoms from the halogen family. Whereas transition metal atoms are selected to replace the Mo atom. The geometric structure of the doped atoms is shown in Fig.\ref{Fig_subst-MoS2}.\\

The Stability of doped MoS$_2$ structures has been  examined by estimating average adsorption energy E$_{ad}$ using the equation
\begin{equation}
E_{ad} = \frac{1}{n}[E_{tot}-E_{MoS_2}-nE_{at}]
\end{equation}
where $E_{tot}$ is the total energy of the functionalized MoS$_2$ unit  cell,  E$_{MoS_2}$ is the total energy of pristine MoS$_2$ in the same unit cell, E$_{at}$ is the per atom energy of the ad-atom and n represents the number of ad-atoms present in the unit cell. 
The adsorption energies for various functionalized MoS$_2$ are listed in the Table.\ref{Table-adsorptionE-MoS2}. 
Relatively large adsorption energies indicate that these atoms can be easily substituted on the pristine MoS$_2$ surface.

\begin{table}[!ht]
	\centering
	\caption{Adsorption energy per ad-atoms adsorbs on MoS$_2$ Monolayer.}
	\begin{tabular}{ c c |c c}
		\hline
		\hline 
		\textbf {ad-atom} & \textbf{ adsorption}& \textbf{ad-atom} & \textbf{ adsorption}\\
		& \textbf{energy(in eV)}&               & \textbf{energy(in eV)}\\ 
		\hline
		N   & -2.139  & F  & -2.039 \\
		As  & -2.001  & Cl & -0.272 \\
		Sb  & -1.208  & Cu & -1.953 \\
		Se  & -2.091  & Ni & -2.142 \\
		\hline
	\end{tabular}
	\label{Table-adsorptionE-MoS2}
\end{table}

\subsection*{Substitution of Sulfur by ad-atoms of group-V and group-VII}
\begin{figure} [t!]
 	\centering
 	\includegraphics[width=1.0\linewidth]{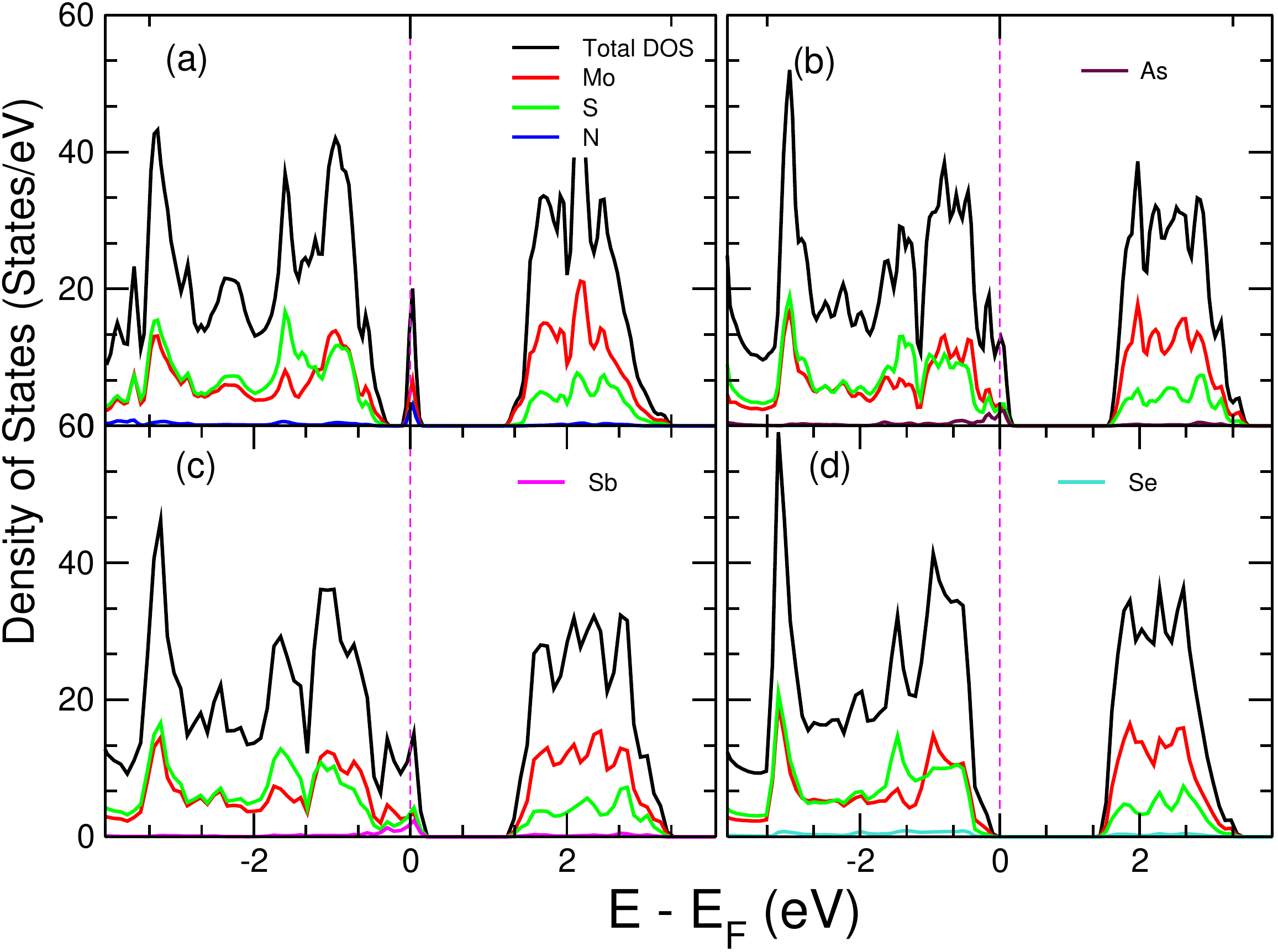}
 	\caption{(color online) Atom projected density of states for functionalized MoS$_2$ with (a)N, (b)As, (c)Sb and (d)Se atoms. Colored curve represents dos from the doped atom. Vertical magenta dashed line is the Fermi energy set at E=0.}
 	\label{Fig_PDOS-MoS2-Group5}
 \end{figure}

We first investigated the effect of substitutional doping of sulfur in the MoS$_2$.   N, As, Sb, and Se from the group-V, F, and Cl from group-VII are considered to replace on S atom in the unit cell.
The atom projected density of states for S atom substituted with group-V ad-atoms are shown in Fig.\ref{Fig_PDOS-MoS2-Group5}. 
Note that the accumulation of energy states from the substituted dopant atoms near the Fermi level occurs in each case accept Se doping.
Since the quantum capacitance is directly proportional to the measure of DOS near the Fermi level, the quantum capacitance is expected to be changed in these systems. The calculated quantum capacitance of the functionalized MoS$_2$ for various group five ad-atom substitution for sulfur is shown in Table.\ref{Table-QC-group5}. Maximum quantum capacitance, 203 $\mu$F/cm$^2$ has been obtained for nitrogen substituted MoS$_2$ monolayer. Whereas no significant change in the electronic structure of MoS$_2$ as well as in quantum capacitance has been observed for Se doping.  
 
 \begin{table}[!ht]
 	\centering
 	\caption{Details of C$_Q$ value calculated  at Fermi energy for various ad-atom functionalized MoS$_2$ Monolayer.}
 	\begin{tabular}{ c c |c c}
 		\hline
 		\hline 
 		\textbf {Configuration} & \textbf{ C$_Q$ }& \textbf{Configuration} & \textbf{C$_Q$ }\\
 		& \textbf{($\mu$F/cm$^2)$}&               & \textbf{($\mu$F/cm$^2)$}\\ 
 		\hline
 		FG - N   &  203.047 &  FG - Sb  & 188.955 \\
 		FG - As  & 189.672  &  FG - Se  & 0.595 	\\	
 		\hline
 	\end{tabular}
 	\label{Table-QC-group5}
 \end{table}
 
\begin{figure} [t!]
 	\centering
 	\includegraphics[width=1\linewidth]{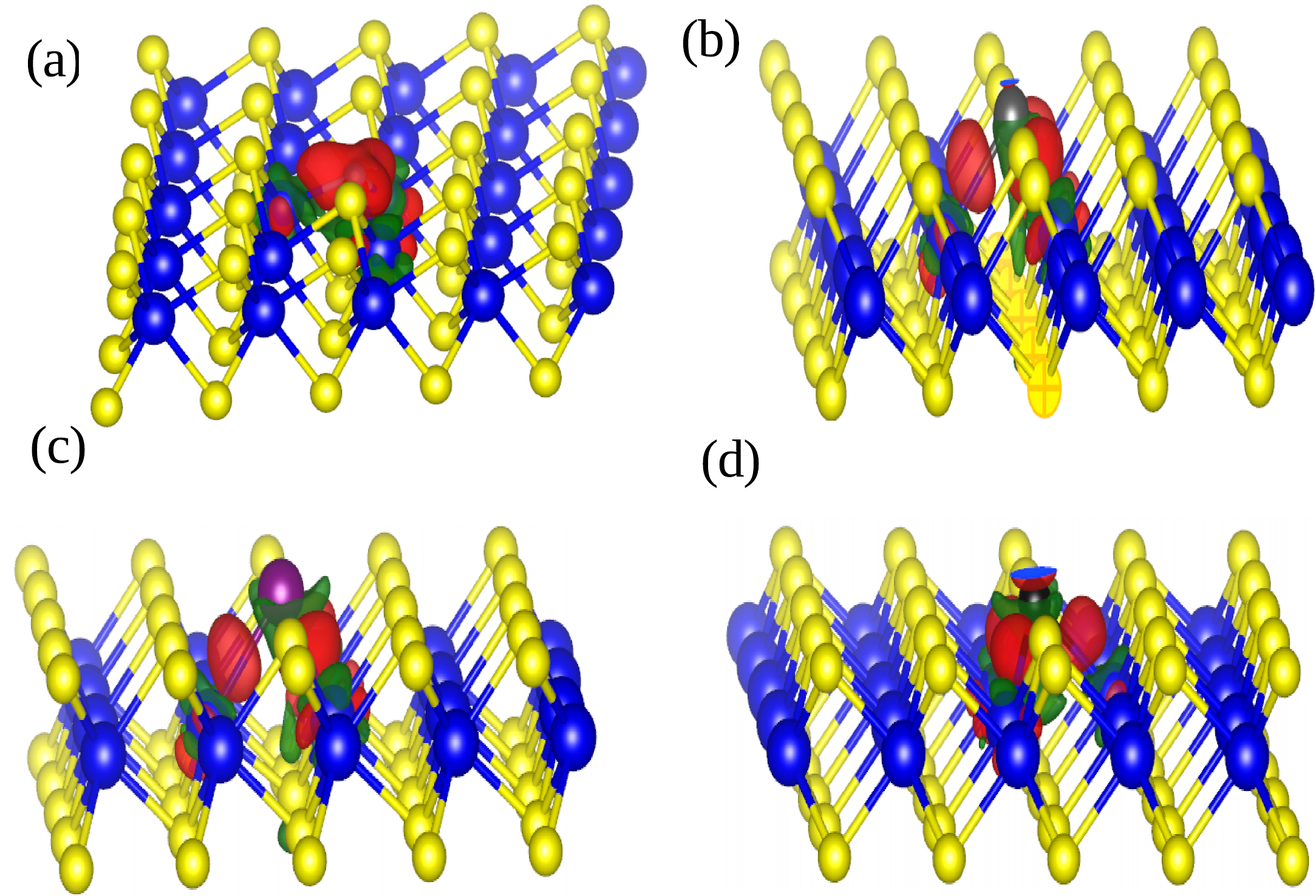}
 	\caption{(color online) Change in  electron density associated with Functionalized MoS$_2$ Monolayer with (a)N, (b)As, (c)Sb and (d)Se doping. Red and green isosurface represnts the charge accumulation and electron deficiency in the system.  The blue and yellow balls represents Mo and S atoms in MoS$_2$ Monolayer.}
 	\label{Fig_MoS2-G5-Iso.eps}
 \end{figure}

 To analyze the result one should note that the N, As, and Sb substitution of S atom on MoS$_2$ monolayer leads to a hole doping condition where the Fermi-level comes close to the valence band. The electron deficiency in the system forces a   charge redistribution in the presence of these three substituted ad-atom, which intern helps to accumulate a large amount of density of states near the valence band edge close to the Fermi energy. To confirm this situation we have calculated the charge density difference, associated with functionalized MoS$_2$ where the average charge density of MoS$_2$ was subtracted from the functionalized charge density. Fig.\ref{Fig_MoS2-G5-Iso.eps} shows the resulted charge distribution upon substitutional doping of MoS$_2$ with N, As, Sb, and Se ad-atoms. The homogeneity of the charge distribution gets significantly disrupted because of the substituted ad-atoms, and charge accumulation near the Mo atom reflects the accumulation of DOS near the valence band edge of the functionalized systems.

\begin{figure} [t!]
 	\centering
 	\includegraphics[width=1.0\linewidth]{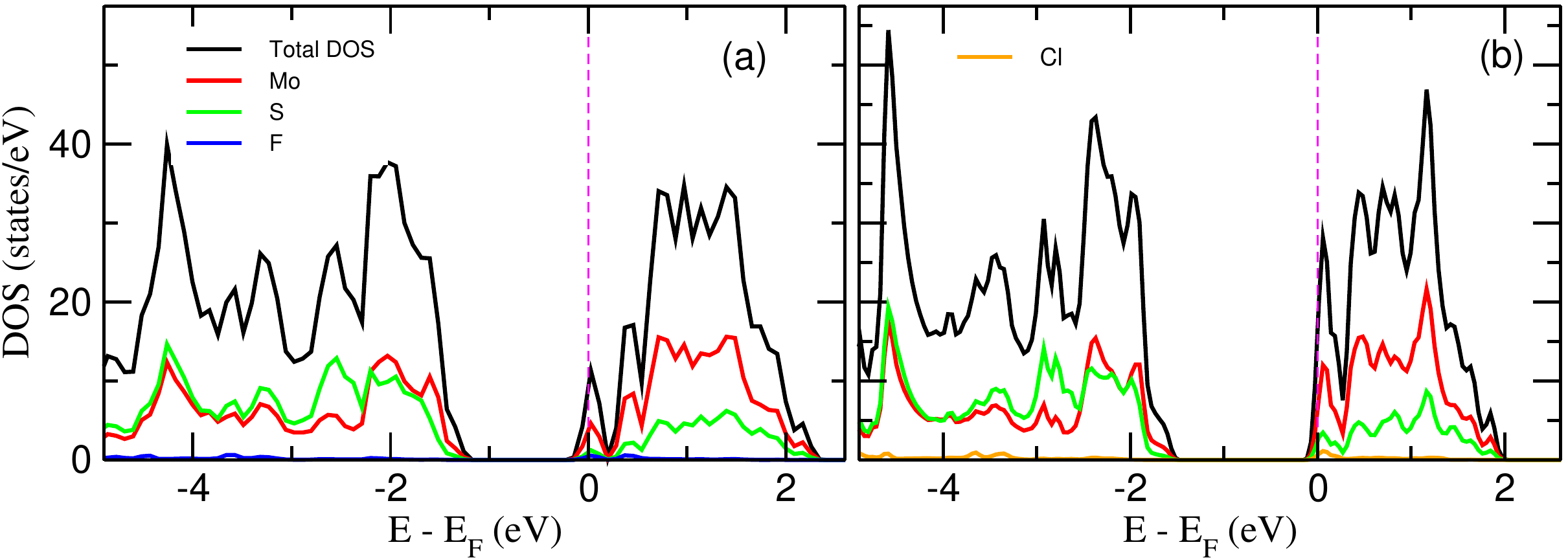}
 	\caption{(color online) Atom projected density of states for functionalized MoS$_2$ with (a)F and (b)Cl atoms. Colored curve represents dos from the doped atom. Vertical majenta dashed line is the Fermi energy set at E=0.}
 	\label{Fig_PDOS-MoS2-Group7.eps}
 \end{figure}

\begin{figure} [t!]
 	\centering
 	\includegraphics[width=1.0\linewidth]{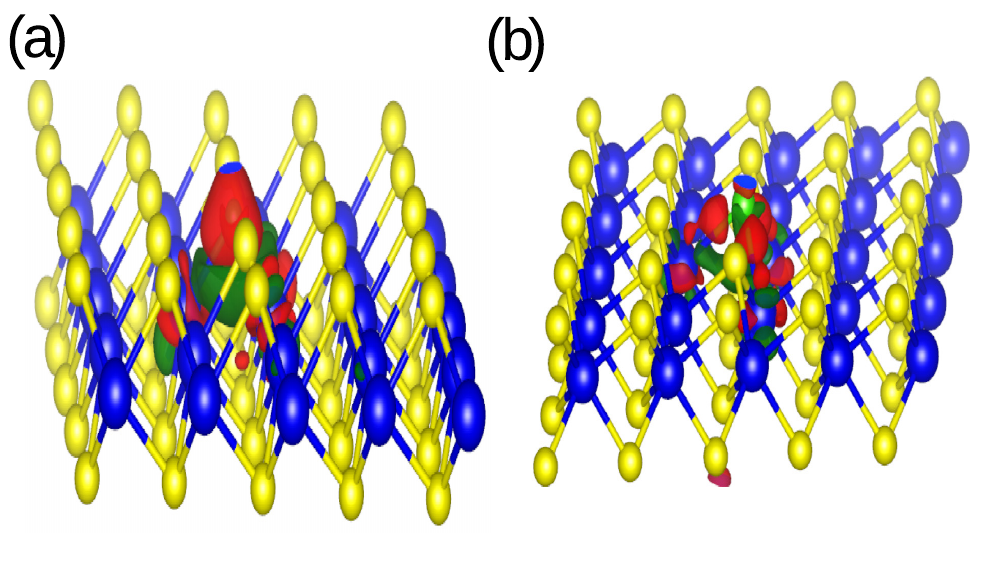}
 	\caption{(color online) Isosurface plots for electron density associated with Functionalized MoS$_2$ Monolayer with (a)F and (b)Cl. Red and green isosurface represnts the charge accumulation and electron deficiency in the system. The blue and yellow balls represents Mo and S atoms in MoS$_2$ Monolayer.}
 	\label{Fig_MoS2-G7-Iso.eps}
 \end{figure}

 In the case of group-VII ad-atoms such as F and Cl functionalization, the substitution of S with these atoms brings the electron doping situation in the system. Here the Fermi energy shifted close to the conduction band and one excess electron in the unit cell will also assist the charge redistribution in the system. Therefore these ad-atoms are capable of significant changes in the DOS as shown in  Fig.\ref{Fig_PDOS-MoS2-Group7.eps}. 
A similar electron density difference plot for F and Cl doped system is shown in  Fig.\ref{Fig_MoS2-G7-Iso.eps} indicates that the electrons are accumulating near the doped site. Further, accumulation of large density of state near the conduction band has also been observed in the partial density of state plot shown in Fig.\ref{Fig_PDOS-MoS2-Group7.eps}. 

\begin{figure} [t!]
 	\centering
 	\includegraphics[width=.9\linewidth]{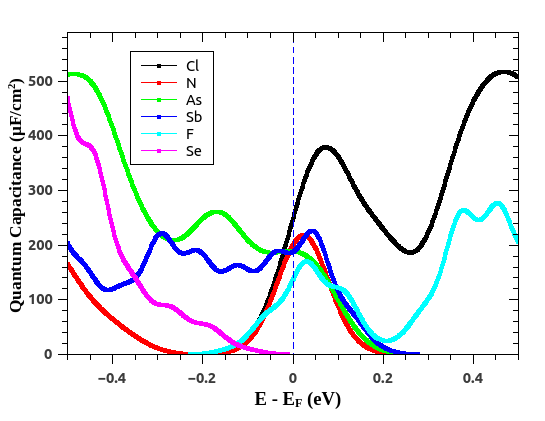}
 	\caption{(color online) The different color curve represents the energy variations of quantum capacitance for MoS$_2$ when S is substituted with various ad-atoms.}
 	\label{Fig-MoS2-Sulfur-Substituted-QC}
 \end{figure}

Accumulation of DOS near the fermi energy enhanced the C$_Q$ value of the functionalized systems. We have calculated 139 $\mu$F/cm$^2$ and 252 $\mu$F/cm$^2$ C$_Q$ values for F and Cl functionalized systems respectively. The energy variation of C$_Q$ in different ad-atom doped(replacement of a S atom) systems are shown in Fig.\ref{Fig-MoS2-Sulfur-Substituted-QC}
 
 \subsection*{Substitution of Mo with Transition Metal(TM) ad-atoms}
\begin{figure} [ht!]
 	\centering
 	\includegraphics[width=1.0\linewidth]{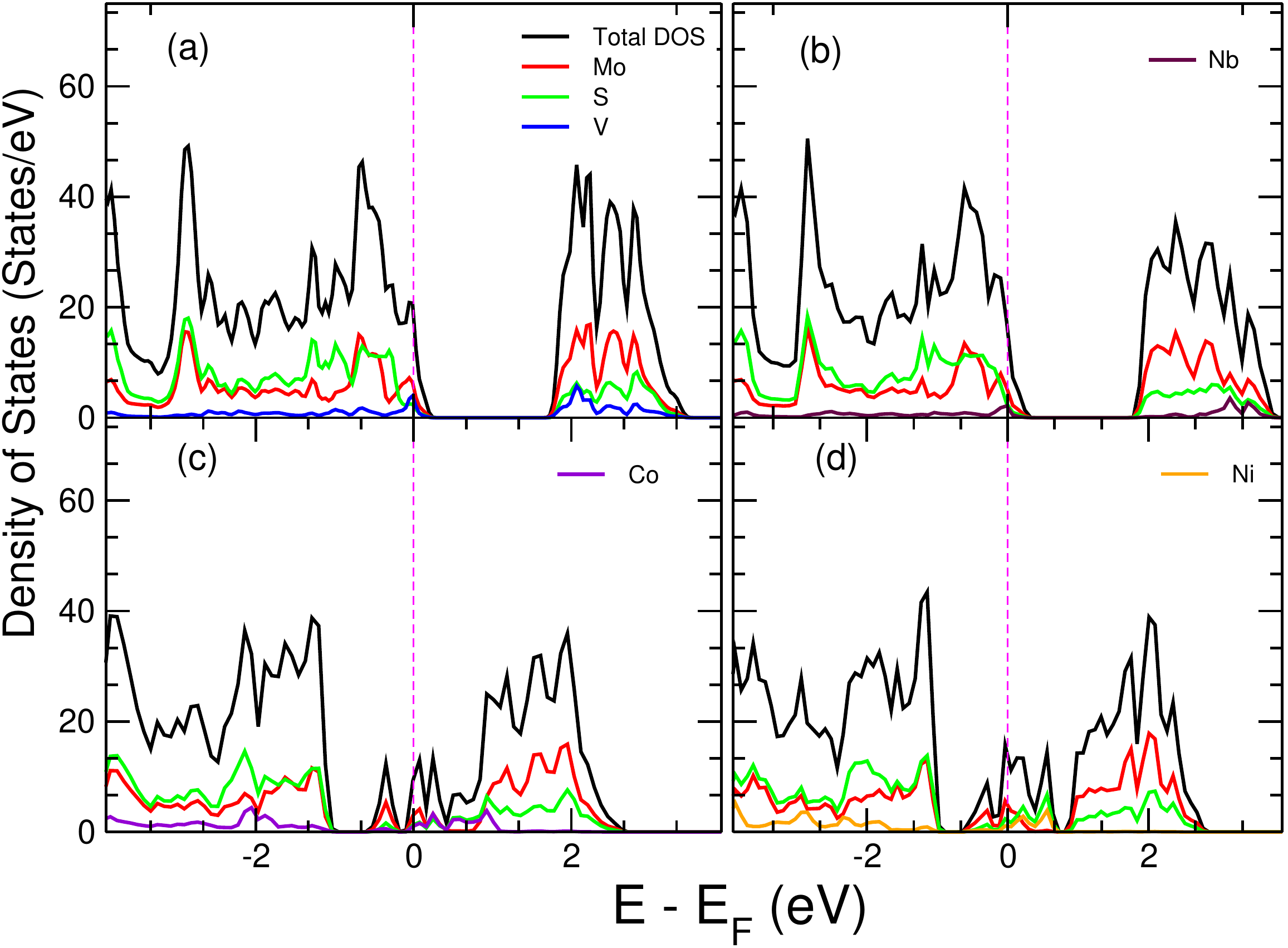}
 	\caption{(color online) Atom projected density of states for functionalized MoS$_2$ with (a)V, (b)Nb, (c)Co and (d)Ni atoms. Colored curve represents dos from the doped atom. Vertical majenta dashed line is the Fermi energy set at E=0.}
 	\label{Fig_PDOS-MoS2-TM.eps}
 \end{figure}

\begin{figure} [t!]
 	\centering
 	\includegraphics[width=0.9\linewidth]{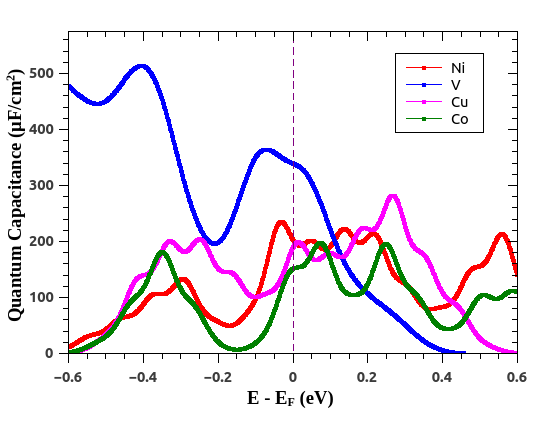}
 	\caption{(color online) The different color curve represents the energy variations of quantum capacitance for MoS$_2$ when Mo is substituted with various ad-atoms.}
 	\label{Fig-MoS2-Mo-Substituted-QC}
 \end{figure}

 Next, we have considered transition metal atoms such as Co, Cu, Ni, and V as ad-atoms to substitute one Mo in the MoS$_2$ monolayer unit cell. 
 The atom projected density of states (Fig.\ref{Fig_PDOS-MoS2-TM.eps}) shows that Co, Cu, Ni, and V are also introduces DOS near Fermi energy.
 Our calculation shows a large change in the C$_Q$ value of upon TM-functionalization of  MoS$_2$ monolayers. The maximum value occurs for V  doped system 263 $\mu$F/cm$^2$, whereas the quantum capacitance for other systems are also high as listed in the  TABLE.\ref{Table-QC-Mo}. The energy variation of C$_Q$ for each TM-doped systems are shown in Fig.\ref{Fig-MoS2-Mo-Substituted-QC}

 \begin{table}[!ht]
 	\centering
 	\caption{Details of C$_Q$ value calculated  at Fermi energy for various ad-atom functionalized MoS$_2$.}
 	\begin{tabular}{ c c |c c}
 		\hline
 		\hline 
 		\textbf {Configuration} & \textbf{ C$_Q$ }& \textbf{Configuration} & \textbf{C$_Q$ }\\
 		& \textbf{($\mu$F/cm$^2)$}&               & \textbf{($\mu$F/cm$^2)$}\\ 
 		\hline
 		FG - Co   &  152.794  &  FG - Ni  & 202.439 \\
 		FG - Cu    &  191.658 &  FG - V   & 263.721 \\	
 		\hline
 	\end{tabular}
 	\label{Table-QC-Mo}
 \end{table}

 \subsection*{Impact of vacancy defects}
 
\begin{figure} [t!]
 	\centering
 	\includegraphics[width=0.5\linewidth]{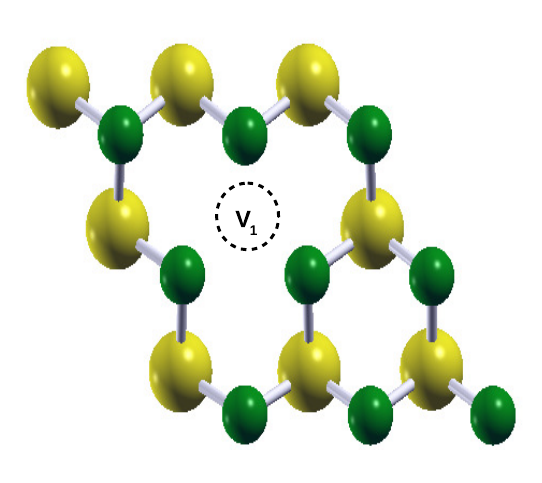}
 	\caption{(color online) Mo-vacancy defected MoS$_2$ structure. Green and yellow ball represents S and Mo atoms respectively.}
 	\label{Fig_defect_str}
 \end{figure}

 Since the presence of vacancy defect can significantly alter the electronic structure of the materials, we, therefore, introduced a vacancy defect in monolayer MoS$_2$. The investigation has been carried out to understand the change in the electronic structure and subsequently the quantum capacitance of the system. Defected monolayer MoS$_2$ systems were modeled by removing a different number of S and Mo atoms from the $3\times 3 $ supercell.  We have considered three different S vacancy concentrations by removing one, two, and three S atoms out of 18 S atoms in the unit cell respectively; ie with 5.5 \%, 11\%, and 16.5\% S vacancy configurations.  In the case of Mo vacancy, we could only investigate 11.5\% vacancy defect configuration, as we found that the structure becomes unstable upon a further increment of the vacancy concentrations.
 Optimized unit cell structure of the Mo vacancy defected structure is shown in Fig.\ref{Fig_defect_str}
  
 \begin{figure} [t!]
 	\centering
 	\includegraphics[width=0.95\linewidth]{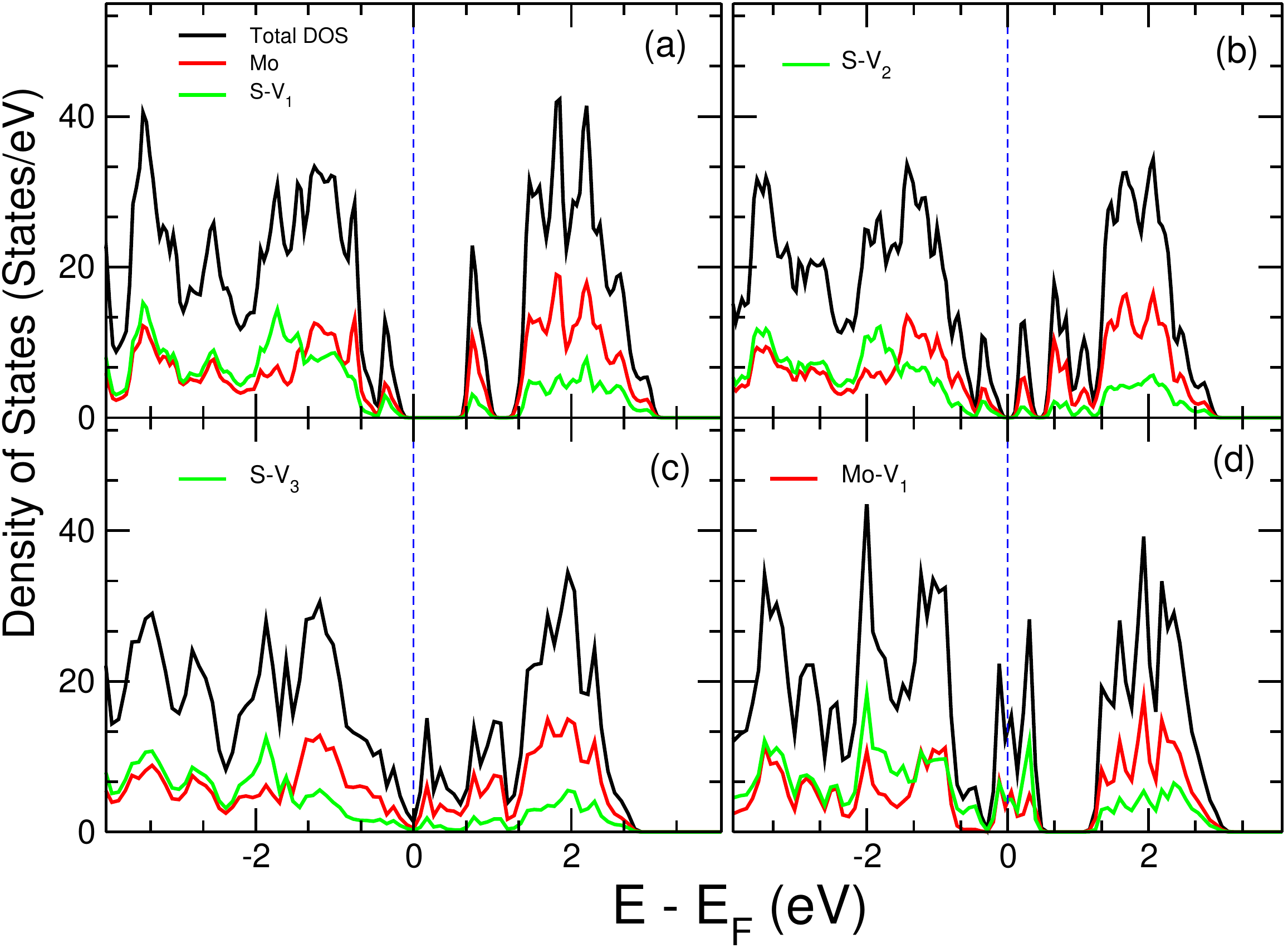}
 	\caption{(color online) Atom projected density of states for MoS$_2$ Monolayer with (a)One sulfur vacancy, (b)two sulfur vacancy, (c)three sulfur vacancy, (d)one Mo vacancy. Vertical blue dashed line is the Fermi energy set at E=0.}
 	\label{Fig_PDOS-MoS2-S-Mo-V1.eps}
 \end{figure}
\begin{figure} [t!]
 	\centering
 	\includegraphics[width=0.9\linewidth]{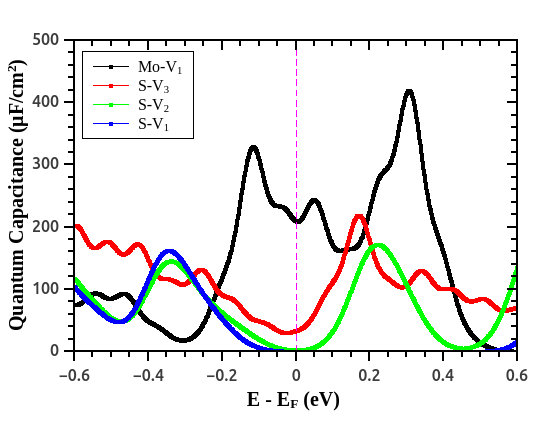}
 	\caption{(color online) The different color curve represents the energy variations of quantum capacitance for MoS$_2$ functionalized with various ad-atoms}
 	\label{Fig-MoS2-Vacancy-QC.eps}
 \end{figure}

 Our calculation shows that the sulfur vacancy in the system has a negligible impact on the value of C$_Q$. The maximum C$_Q$ value obtained is 33 $\mu$F/cm$^2$ for 16.5\% defect configurations. However, in the case of 11\% Mo vacancy defect system show a large quantum capacitance of 209.733 $\mu$F/cm$^2$. The result can be analyzed from the obtained electronic structure of the defective system as shown in Fig.\ref{Fig_PDOS-MoS2-S-Mo-V1.eps}. Mo-vacancy significantly changes the electronic structure and introduced a large amount of DOS near the Fermi energy and the system behaves as conductors. Whereas the change in DOS near the Fermi energy is negligible in the case of S-vacancy systems.
The energy variation of C$_Q$ for different configurations are shown in Fig.\ref{Fig-MoS2-Vacancy-QC.eps} and the calculated quantum capacitances are listed in  Table.\ref{Table-QC-vacancy}. 
 
 \begin{table}[!ht]
 	\centering
 	\caption{Details of C$_Q$ value calculated  at Fermi energy for various ad-atom functionalized MoS$_2$.}
 	\begin{tabular}{ c c }
 		\hline
 		\hline 
 		\textbf {Configuration} & \textbf{ C$_Q$ }\\
 		& \textbf{($\mu$F/cm$^2)$}\\ 
 		\hline
 		MoS$_2$ with 5.5\% S vacancy    &  0.190 \\
 		MoS$_2$ with 11\% S vacancy    &  2.500 \\
 		MoS$_2$ with 16.5\% S vacancy  &  33.665 \\
 		MoS$_2$ with 11.5\% Mo vacancy   &  209.733 \\
 		\hline
 	\end{tabular}
 	\label{Table-QC-vacancy}
 \end{table}

 \section*{Conclusion}
 In conclusion, we have studied the quantum capacitance in functionalized  MoS$_2$ monolayer.  Our theoretical investigation shows that the quantum capacitance (C$_Q$) of MoS$_2$ electrodes can be enhanced significantly by introducing ad-atoms and vacancy defects in the  MoS$_2$ monolayer sheet.  A marked quantum capacitance above 200 $\mu$ F/cm$^2$ has been observed. These calculations show that the quantum capacitance of MoS$_2$ monolayer enhances with substitutional doping of Mo with transition metal adatoms.  A significant charge transfer and charge redistribution occur in ad-atom doped MoS$_2$ monolayer results in an accumulation of a large number of electronic states near Fermi level. The additional charge carriers brought by ad-atoms change the carrier concentration in monolayer MoS$_2$  that leads to the shift of the Fermi level and significantly improves the quantum capacitance of the system.
\section*{References}
\bibliographystyle{iopart-num}
\bibliography{ref}
\end{document}